\documentclass[twocolumn]{aastex61}
\pdfoutput=1 
\usepackage{amsmath,amstext}
\usepackage[T1]{fontenc}
\usepackage{apjfonts} 
\usepackage[figure,figure*]{hypcap}
\usepackage[hang,flushmargin]{footmisc} 
\usepackage{lineno}

\usepackage[caption=false]{subfig}
\usepackage{graphicx}

\graphicspath{{./figs/}}

\usepackage{hyperref}
\usepackage{url}

\usepackage{enumitem}
\setlist[enumerate]{label*=\arabic*.}

\newcommand{\apx}{\ensuremath{\sim}}

\shorttitle{Kilonova Contamination}
\shortauthors{Cowperthwaite et~al.}

\begin{document}

\title{An Empirical Study of Contamination in Deep, Rapid, and \\ Wide-Field Optical Follow-Up of Gravitational Wave Events}

\author{P.~S.~Cowperthwaite}
\affiliation{Harvard-Smithsonian Center for Astrophysics, 60 Garden Street, Cambridge, Massachusetts 02138, USA}
\affiliation{NSF GRFP Fellow, e-mail: pcowpert@cfa.harvard.edu}
\author{E.~Berger}
\affiliation{Harvard-Smithsonian Center for Astrophysics, 60 Garden Street, Cambridge, Massachusetts 02138, USA}
\author{A.~Rest}
\affiliation{Space Telescope Science Institute, 3700 San Martin Drive, Baltimore, MD 21218, USA}
\affiliation{Department of Physics and Astronomy, The Johns Hopkins University, 3400 North
Charles Street, Baltimore, MD 21218, USA}
\author{R.~Chornock}
\affiliation{Astrophysical Institute, Department of Physics and Astronomy, 251B Clippinger Lab, Ohio University, Athens, OH 45701, USA}
\author{D.~M.~Scolnic}
\affiliation{Kavli Institute for Cosmological Physics, The University of Chicago, Chicago, IL 60637}
\author{P.~K.~G.~Williams}
\affiliation{Harvard-Smithsonian Center for Astrophysics, 60 Garden Street, Cambridge, Massachusetts 02138, USA}
\author{W.~Fong}
\affiliation{CIERA and Department of Physics and Astronomy, Northwestern University, Evanston, IL 60208}
\author{M.~R.~Drout}
\affiliation{The Observatories of the Carnegie Institution for Science, 813 Santa Barbara St., Pasadena, CA 91101}
\affiliation{Hubble, Carnegie-Dunlap Fellow}
\author{R.~J.~Foley}
\affiliation{Department of Astronomy and Astrophysics, University of California, Santa Cruz, CA 95064, USA}
\author{R.~Margutti}
\affiliation{CIERA and Department of Physics and Astronomy, Northwestern University, Evanston, IL 60208}
\author{R.~Lunnan}
\affiliation{The Oskar Klein Centre \& Department of Astronomy, Stockholm University, AlbaNova, SE-106 91 Stockholm, Sweden}
\author{B.~D.~Metzger}
\affiliation{Department of Physics and Columbia Astrophysics Laboratory, Columbia University, New York, NY 10027, USA}
\author{E.~Quataert}
\affil{Department of Astronomy \& Theoretical Astrophysics Center, University of California, Berkeley, CA 94720-3411, USA}

\begin{abstract} 
We present an empirical study of contamination in deep, rapid, and wide-field optical follow-up searches of gravitational wave sources from Advanced LIGO/Virgo (ALV). We utilize dedicated observations during four nights of imaging with the Dark Energy Camera (DECam) wide-field imager on the Blanco 4-m telescope at CTIO. Our search covered \apx56 deg$^2$, with two visits per night separated by $\approx 3$~hours, in $i$- and $z$-band, followed by an additional set of $griz$ images three weeks later to serve as reference images for subtraction, and for the purpose of identifying galaxy and stellar counterparts for any transient sources. We achieve $5\sigma$ point-source limiting magnitudes of $i \approx 23.5$ and $z \approx 22.4$ mag in the coadded single-epoch images. We conduct a search for transient objects that can mimic the $i-z$ color behavior of both red ($i-z > 0.5$~mag) and blue ($i-z < 0$~mag) kilonova emission, finding 11 and 10 contaminants, respectively. Independent of color, we identify 48 transients of interest. Additionally, we leverage the rapid cadence of our observations to search for sources with characteristic timescales of $\approx1$ day and $\approx3$ hours, finding no potential contaminants. We assess the efficiency of our pipeline and search methodology with injected point sources, finding that we are 90\% (60\%) efficient when searching for red (blue) kilonova-like sources to a limiting magnitude of $i \lesssim 22.5$ mag. Applying these efficiencies, we derive sky rates for kilonova contaminants in the red and blue regimes of $\mathcal{R}_{\rm red} \approx 0.16$ deg$^{-2}$ and $\mathcal{R}_{\rm blue} \approx 0.80$ deg$^{-2}$. The total contamination rate, independent of color, is $\mathcal{R}_{\rm all} \approx 1.79$ deg$^{-2}$. We compare our derived results to optical follow-up searches of the gravitational wave events GW150914 and GW151226 and comment on the outlook for GW follow-up searches as additional GW detectors (e.g., KAGRA, LIGO India) come online in the next decade. 
\end{abstract}


\section{Introduction}
\label{sec:intro}
The detection of gravitational wave (GW) events during the first and second observing runs of the Advanced Laser Interferometer Gravitational-Wave Observatory (aLIGO) ushered in the era of GW astronomy. The three announced detections to date (GW150914, GW151226, GW170104, and GW170814) were due to binary black hole (BBH) mergers, with component masses of $\approx8-36$ M$_\odot$ \citep{gw150914,gw151226,gw170104,gw170814}. Yet to be detected are the mergers of compact binaries containing at least one neutron star (NS) such as a binary neutron star (BNS) or neutron star-black hole (NS-BH) system. In such mergers we expect electromagnetic (EM) counterparts, including short gamma-ray bursts and their broad-band afterglows (SGRB, see e.g., \citealt{fong13, berger14, fong+15}) and kilonovae, isotropic thermal optical/NIR transients powered by the radioactive decay of $r$-process nuclei synthesized in the merger ejecta \citep[see e.g.,][]{kasen+13,barnes13,metzger17}. On longer timescales (e.g., weeks-years), the dynamical interaction of the merger ejecta with the ambient medium can produce radio emission \citep{nakar12}.

While the GW data offer an unprecedented view of the dynamics of these systems and allows for tests of general relativity in a strong gravity regime not previous accessible, the full science potential of these detections cannot be realized in the absence of detected EM counterparts. An EM counterpart will allow a unique association with a host galaxy, a precise distance measurement, insight into the system's local environment, and a probe of ejecta hydrodynamics following the merger \citep[e.g.,][]{metzger12}. 
 
In the context of searches for these EM counterparts, the wide-field instruments necessary to cover the large localization regions are most effective in gamma-rays (but emission in this band is strongly beamed) and optical wavelengths. For wide-field searches conducted in the optical bands, kilonovae have been of particular interest due to the isotropic nature of the emission and the expectation that this emission should accompany all BNS mergers, and some of the  NS-BH mergers in which the NS is disrupted outside the event horizon \citep{metzger12}. The expected short duration ($\lesssim {\rm week}$) and low luminosity ($L_{\rm bol} \sim 10^{41}$~erg s$^{-1}$ at peak) of the kilonova emission make deep and rapid searches imperative. Coupled with the large search regions, this also requires the development of methodologies to robustly identify kilonovae among the vast numbers of potential contaminating sources. For example, optical follow-up of GW151226, covering hundreds of square degrees, has led to the identification of tens of unrelated transient sources despite being shallower than ideal searches for kilonova emission require \citep[e.g.,][]{cenko15,cenko16,smartt+16b}.

In \citet[hereafter CB15]{cowp15}, we addressed the question of kilonova detectability and the associated contamination with a Monte Carlo method to simulate tens of thousands of observations of both kilonovae and potential contaminating sources (e.g., supernovae and other known and speculative rapid transients), exploring a variety of cadences and search depths. We found that nightly observations to $5\sigma$ limiting magnitudes of $i \approx 24$ and $z \approx 23$ are required to achieve a 95\% kilonova detection rate. Furthermore, the simulated observations revealed that kilonovae occupy a unique region of phase-space defined by $i-z$ color and rise time $(t_{\rm rise})$. The analysis suggests that kilonovae can be identified among contaminating sources using $i-z\gtrsim 0.3$ mag and $t_{\rm rise}\lesssim 4$ d. This motivates the need for a new and empirical investigation into the issue of contamination.

In this paper, we extend our work in CB15 with an empirical investigation of contamination using a set of deep, rapid-cadence observations obtained for this specific purpose with the Dark Energy Camera (DECam; \citealt{flaugher+15}). These observations encompass four nights of data, covering 56 deg$^2$, in $i$- and $z$-bands, with two visits per night separated by $\approx 3$ hr, and a depth designed to match expected kilonova brightness within aLIGO's sensitivity volume. This dataset is unique as it targets cadences shorter than those employed by typical time-domain surveys or by any of the existing follow-up observations of aLIGO BBH merger detections, allowing direct insight into contamination in rapid searches. We use these data to conduct a search for transient sources that could mimic the $i-z$ color behavior of kilonova, considering both red ($i-z \gtrsim 0.5$~mag) and blue ($i-z \lesssim 0$~mag) scenarios.

The paper is organized as follows. In \autoref{sec:kn_models} we describe the various kilonova models explored in this work. We present the observations and data analysis in \autoref{sec:obs}. In \autoref{sec:search} we discuss the methodologies and results of our search for kilonova contaminants. In \autoref{sec:fakes}, we determine our search efficiency using point sources injected into the images with a wide range of brightness, fading rate, and color. In \autoref{sec:contam_rates} we determine the contamination rate as a function of $i-z$ color and compare our results to those reported from optical follow-up of the current BBH merger events. We discuss the implications of our search in the context of current and future GW detectors and follow-up facilities in \autoref{sec:conc}.

All magnitudes presented in this work are given in the AB system unless otherwise noted. Cosmological calculations are performed using the cosmological parameters $H_0 = 67.7$ km s$^{-1}$ Mpc$^{-1}$, $\Omega_M = 0.307$, and $\Omega_{\Lambda} = 0.691$ \citep{planck15}.

\section{Overview of Kilonova Models}
\label{sec:kn_models}
The mergers of compact binaries containing at least one NS are expected to produce ejecta through dynamical processes such as tidal forces and accretion disk winds \citep{goriely+11,bauswein+13,fernandez+15,radice+16, metzger17}. Numerical simulations indicate that the unbound debris has a mass of $M_{\rm ej} \sim 10^{-4} - 10^{-1}$~M$_{\odot}$ with a velocity of $\beta_{\rm ej} \sim 0.1 - 0.3$, with a dependence on parameters such as the mass ratio and equation of state \citep{hotokezaka+13, tanaka+14, kyutoku+15}. The ejecta are expected to be neutron rich, with a typical electron fraction of $Y_e \lesssim 0.3$, with simulations showing a range of values from $Y_e \sim 0.1 - 0.3$ \citep{goriely+11, bauswein+13, sekiguchi+15, radice+16,siegel17}. This electron fraction is low enough ($Y_e \lesssim 0.25$) that the ejecta are expected to undergo $r$-process nucleosynthesis, producing heavy elements $(A \gtrsim 130)$, particularly in the lanthanide and actinide groups \citep{goriely+11,bauswein+13,siegel17}. These groups of elements have open f-shells which allow a large number of possible electron configurations, resulting in a large opacity in the bluer optical bands ($\kappa_{\nu} \sim$ 100 cm$^2$ g$^{-1}$ for $\lambda \sim 1$~$\mu$m, see e.g. Figure 10 in, \citealt{kasen+13}). A more recent calculation by \cite{fontes+17} suggests that the lanthanide opacities may be an order of magnitude higher ($\kappa_{\nu} \sim 1000$  cm$^2$ g$^{-1}$ for $\lambda \sim 1$~$\mu$m).

Radioactive decay of the $r$-process elements synthesized during the merger heats the ejecta producing an isotropic, thermal transient \citep{lp98, rosswog05, metzger10,tanaka+14}. The combination of low ejecta mass and high ejecta velocity, coupled with the strong optical line blanketing, results in a transient that is faint ($i \approx 23$ and $z \approx 22$ mag at 200 Mpc), red ($i-z \gtrsim 0.5$ mag), and short-lived with a typical duration of $\sim {\rm few}$ days in $z$-band and $\sim{\rm week}$ in $J$-band \citep{barnes13,barnes16,metzger17}. In the case of larger opacities \citep[e.g.,][]{fontes+17}, the transient is expected to peak in the IR (\apx3~$\mu$m), with a duration of \apx10~days \citep{fontes+17,wollaeger+17}.

In addition to the neutron-rich dynamical ejecta, recent work has suggested that the mergers can also produce ejecta with a high electron fraction $(Y_e > 0.25$,  \citealt{wanajo+14,goriely+15}) if BNS mergers lead to a hypermassive neutron star \citep[HMNS; see e.g.,][]{sekiguchi+11} with a lifetime of $\gtrsim 100$ ms. The resulting HMNS irradiates the disk wind ejecta with a high neutrino luminosity which raises the electron fraction of the material and suppresses $r$-process nucleosynthesis\footnote{Even if a HMNS star is not formed, a sufficiently rapidly spinning black hole may lead to a small amount of ejecta with $Y_e \gtrsim 0.25$ \citep[see e.g.,][]{fernandez13,fernandez+15}. In this scenario, the resulting kilonova emission is broadly identical to the HMNS case, but the lower ejecta mass leads to a faster and fainter transient.}. This material will have an opacity similar to the Fe-peak opacities seen in Type Ia SNe, producing emission that is slightly brighter ($r \approx 22$ mag at 200 Mpc), bluer ($i-z \lesssim 0$), and shorter-lived ($\approx 1-2$ days, \citealt{metzger14,kasen+15}). This blue kilonova component has a strong dependence on viewing angle, with the polar regions of the ejecta being exposed to the highest neutrino flux \citep{metzger14,kasen+15}. Consequently, if the merger is viewed face-on $(\theta \approx$ 15--30 deg), this blue component may be visible. We expect that up to half of mergers will be viewed at such angles \citep[see e.g.,][]{metzger12}. However, at larger viewing angles, the lanthanide-rich material in the dynamical ejecta will obscure the blue emission, and only a red kilonova will be observed \citep{kasen+15,metzger17}. While this component will be brighter than the expected $r$-band emission from the lanthanide-rich material, its detection requires rapid-cadence observations within a few hours of the GW detection (CB15).

Lastly, it has been argued that a small fraction of the merger ejecta may expand so rapidly that it is unable to undergo $r$-process nucleosynthesis \citep{bauswein+13}. This material instead deposits energy into the ejecta via neutron $\beta$-decay. At very early times $(\lesssim 1$ hr post merger), the specific heating rate from the neutron $\beta$-decay is an order of magnitude higher than that generated by the $r$-process nuclei \citep{metzger17}. This timescale is well matched to the diffusion time for the free neutron ejecta resulting in bright brief emission. For an ejecta mass with $10^{-4}$ M$_{\odot}$ of free neutrons, the resulting transient will have a peak $r$-band magnitude of $\approx 22$ mag at 200 Mpc with a characteristic timescale of $\sim 1-2$ hours \citep{metzger15}. This speculative early time emission is often referred to as a ``neutron precursor." However, due to the high velocity of the free neutrons $>0.4$c, this component of the ejecta may be visible before the equatorial lanthanide-rich ejecta and thus be observable for a wider range of viewing angles than the blue kilonova. This transient will be as bright as the blue kilonova emission but significantly shorter in duration requiring particularly rapid observations in response to a GW trigger (CB15).

To summarize, kilonova emission with red color $(i-z\gtrsim 0.5)$, a peak brightness of $z\approx22.2$~mag at 200 Mpc, and a duration of $\sim \rm few$ days is expected to be ubiquitous.  Blue kilonova emission due to a surviving HMNS is expected to be bluer ($i-z \lesssim 0$~mag), similarly faint ($r \approx 22$~mag) and shorter in duration ($\lesssim 1-2$ days). The predicted behavior and composition of the dynamical ejecta and associated red kilonova emission is robust and ubiquitous, however the nature and observability of the blue kilonova emission depends both on the fraction of cases in which a HMNS survives (unknown) and on geometrical effects ($\lesssim 50\%$).  Finally, emission due to free neutrons will be similar in brightness and color to the blue kilonova emission, but with a timescale of only $\sim \rm few$ hours; the prevalence of this signal is uncertain.  The observations described in this paper address contamination in all of these cases.

\section{Observations and Data Analysis} 
\label{sec:obs}

We obtained data for this study using the DECam imager on the Blanco 4-m telescope at the Cerro Tololo Inter-American Observatory (CTIO)\footnote{PI: Berger, \href{https://www.noao.edu/perl/abstract?2013A-0214}{NOAO 2013A-0214}}. DECam is a wide-field optical imager with a 3.3 deg$^2$ field-of-view and a CCD sensitive out to $\sim 1\mu$m \citep{flaugher+15}, making it an ideal instrument for optical follow-up covering the sizable GW localization regions, particularly in the context of red kilonova emission. The dataset consists of 21 contiguous pointings, covering $\approx56$~deg$^2$ in the Antlia cluster,\footnote{The effective area corresponds to 21 DECam pointings accounting for an overall $\approx 20\%$ loss of area due to chip gaps (10\%), three unused CCDs (5\%, see \autoref{sec:data_cuts} and \citealt{diehl+14}), and masked edge pixels (5\%).} with observations conducted nightly over a five day period (2013 March $1-5$ UT; ``search images''); however due to poor weather conditions the data taken on 2013 March 4 UT were unusable. Two sets of observations, separated by $\approx 3$ hours, were obtained during each observing night, with each observation consisting of two 150~s exposures in $i$-band and two 60~s exposures in $z$-band. We obtained an additional epoch for image subtraction on 2013 March 22 UT. This epoch consists of observations in $griz$-bands to determine colors for any template counterparts, with two 85~s exposures in $g$- and $r$-band, and the same exposures in $i$- and $z$-band as the initial search images. We list the central pointing coordinates for the 21 fields in \autoref{tab:pointings}, and summarize the observations in Tables~\ref{tab:search} and~\ref{tab:templ}. 

We processed the data using {\tt photpipe}, an image processing pipeline used by several previous time domain surveys (see e.g., \citealt{rest+05,rest+14}) to perform single-epoch image processing, image subtraction, and candidate identification. The single epoch processing steps include initial image reduction (e.g., bias subtraction, cross-talk corrections, flat-fielding), astrometric and photometric calibration, and coaddition of individual pairs of exposures. Search and template images are deprojected into a tangential plane using {\tt SWARP} \citep{bertin+02}. Difference imaging is performed in {\tt photpipe} using the {\tt hotpants} software package \citep{alard00,becker15}. Source detection and point spread function (PSF) photometry is performed on the subtracted images using an implementation of {\tt DoPhot} (\citealt{schechter+93}) that has been optimized for difference images. 

We start with the raw images and appropriate calibration files obtained from the NOAO archive\footnote{\url{http://archive.noao.edu/}}. We performed astrometric calibration relative to the 2MASS $J$-band point-source catalog. We then coadded the pairs of $i$- and $z$- band images from a single epoch. We performed photometric calibration using the PS1 3$\pi$ survey to compute zeropoints for SDSS Stripe~82 standard images taken on each observing night. We applied appropriate corrections between PS1 and DECam magnitudes to these zeropoints \citep{scolnic+15}. We propagated the corrected zeropoints to the science observations with appropriate scaling for exposure time and airmass. We then performed image subtraction using the 2013 March 22 UT observations as template images. 

\begin{deluxetable}{lcc}
\tabletypesize{\footnotesize}
\tablecolumns{3}
\tablewidth{0pt}
\captionsetup{justification=centering}
\tablecaption{Field Pointing Centers
\label{tab:pointings}}	
\tablehead{
    \colhead{Pointing} &
    \colhead{R.A.} &  
    \colhead{Decl.} \\}
\startdata
Antlia 1 & 10:19:58.5  & $-$30:55:24 \\
Antlia 2 & 10:19:44.5  & $-$33:07:24 \\
Antlia 3 & 10:19:30.5  & $-$35:19:24 \\
Antlia 4 & 10:19:10.5  & $-$37:31:24 \\
Antlia 5 & 10:30:03.5  & $-$30:55:24 \\
Antlia 6 & 10:30:03.5  & $-$33:07:24 \\
Antlia 7 & 10:30:03.5  & $-$35:19:24 \\
Antlia 8 & 10:30:03.5  & $-$37:31:24 \\
Antlia 9 & 10:40:08.5  & $-$30:55:24 \\
Antlia 10 & 10:40:22.5  & $-$33:07:24 \\
Antlia 11 & 10:40:38.5  & $-$35:19:24 \\
Antlia 12 & 10:40:57.5  & $-$37:31:24 \\
Antlia 13 & 10:50:12.5  & $-$30:55:24 \\
Antlia 14 & 10:50:41.5  & $-$33:07:24 \\
Antlia 15 & 10:51:13.5  & $-$35:19:24 \\
Antlia 16 & 10:51:50.5  & $-$37:31:24 \\
Antlia 17 & 11:00:17.5  & $-$30:55:24 \\
Antlia 18 & 11:01:00.5  & $-$33:07:24 \\
Antlia 19 & 11:01:48.5  & $-$35:19:24 \\
Antlia 20 & 11:02:44.5  & $-$37:31:24 \\
Antlia 21 & 10:36:36.0  & $-$27:31:04 \\
\enddata
\tablecomments{Central J2000 coordinates for all 21 fields used in this analysis.}
\end{deluxetable}

\begin{deluxetable*}{lcrcccccc}
\tabletypesize{\footnotesize}
\tablecolumns{9}
\tablewidth{0pt}
\tablecaption{Summary of Search Observations
\label{tab:search}}	
\tablehead{
    \colhead{Night}    &
    \colhead{Epoch} &  
    \colhead{UT}       & 
    \colhead{$\langle$PSF$_i$$\rangle$}    &
    \colhead{$\langle$PSF$_z$$\rangle$}    &
    \colhead{$\langle$airmass$\rangle$}  &
    \colhead{$\langle$depth$_i\rangle$}   &
    \colhead{$\langle$depth$_z\rangle$} \\
    \colhead{ } &
    \colhead{ } & 
    \colhead{ } &
    \colhead{(arcsec)} &
    \colhead{(arcsec)} &
    \colhead{ } &
    \colhead{(mag)} &
    \colhead{(mag)}
}
\startdata
	1 & A & 2013-03-01 & 0.92 & 0.86 & 1.20 & 23.2 & 22.3 \\  
           & B & 2013-03-01 & 1.01 & 0.97 & 1.03 & 23.2 & 22.4 \\
  	2 & A & 2013-03-02 & 1.08 & 1.05 & 1.22 & 23.2 & 22.1 \\ 
           & B & 2013-03-02 & 1.09 & 1.06 & 1.03 & 23.4 & 22.3 \\
  	3 & A & 2013-03-03 & 0.90 & 0.87 & 1.13 & 23.6 & 22.4 \\ 
            & B & 2013-03-03 & 0.91 & 0.87 & 1.03 & 23.6 & 22.5 \\
  	4 & A & 2013-03-05 & 0.87 & 0.82 & 1.15 & 23.7 & 22.5 \\ 
           & B & 2013-03-05 & 0.92 & 0.89 & 1.07 & 23.7 & 22.6 \\
\enddata
\tablecomments{Summary of our DECam observations used as search epochs. The data taken on 2013 March 4 UT are omitted as they are unusable due to poor weather conditions. The PSF and airmass are averaged across all observations on a given date. The $5\sigma$ point source depth is the mean value computed for the coadded search images.}
\end{deluxetable*}

\begin{deluxetable}{lcccc}
\tabletypesize{\footnotesize}
\tablecolumns{5}
\tablewidth{0pt}
\tablecaption{Summary of Template Observations
\label{tab:templ}}	
\tablehead{
    \colhead{Filter} &
    \colhead{$\langle$PSF$\rangle$}    &
    \colhead{$\langle$airmass$\rangle$}  &
    \colhead{$\langle$depth$\rangle$} \\
    \colhead{ } &
    \colhead{(arcsec)} &
    \colhead{ } &
    \colhead{(mag)}
}
\startdata
$g$-band & 1.03 & 1.07 & 22.8 \\
$r$-band & 0.95 & 1.08 & 23.0 \\
$i$-band & 0.90 & 1.09 & 23.2 \\
$z$-band & 0.87 & 1.09 & 22.1 \\
\enddata
\tablecomments{Summary of our DECam observations used for the
template epoch. All data were taken on 2013 March 22 UT. Values
are computed as for \autoref{tab:search}}
\end{deluxetable}

In addition to the photometry performed by {\tt photpipe}, we also constructed secondary catalogs for all sources identified in the $griz$-band template epochs. This was accomplished using the {\tt Source Extractor} ({\tt SExtractor}) photometry package in single-image mode \citep{bertin96}. We also performed forced aperture photometry in the template epoch at the position of each candidate identified by {\tt photpipe}. This approach helps to identify the presence of flux in the template images for objects not detected by {\tt SExtractor} which is indicative of image artifacts and defects. We use these additional template catalogs as a useful tool for candidate classification and artifact rejection (see \autoref{sec:source_cuts}).

Our search images achieve an average 5$\sigma$ depth of $i \approx 23.5$ and $z \approx 22.4$~mag for point-sources in the coadded single-epoch images (\autoref{tab:search}). The coadded template images achieve a $5\sigma$ depth for point-sources of $g \approx 22.8$, $r \approx 23.0$, $i \approx 23.2$ and $z \approx 22.1$~mag (\autoref{tab:templ}). The estimated average 5$\sigma$ limiting magnitudes for point sources in the difference images are $i \approx 23.2$ and $z \approx 22.3$ mag (see \autoref{sec:fakes}). There is a mean scatter in the $5\sigma$ depths in the search and difference images of $\approx 0.2$~mag between epochs.

\newpage
\section{A Search for Optical Transients}
\label{sec:search}

Our primary goal is to uncover all optical transients in our data and then to determine specifically the areal rate of kilonova contaminants.

\subsection{Selection Based On Data Quality and Fading Behavior}
\label{sec:data_cuts}

Our initial selection criteria are designed to identify transient sources in our search images that have sufficient data quality for further analysis. These selection criteria are:

\begin{enumerate}
\item Given that we are simulating follow-up triggered by a GW detection, we search for sources that are detected at the beginning of our search (i.e., we treat our first night of observations as if it followed a GW detection notice). We accomplish this by requiring four $5\sigma$ detections in any combination of $i$- or $z$-band across the first four epochs (i.e., four total detections across the first two nights).  This selection criterion leads to an initial sample size of 2818 sources.

\item We expect both kilonovae and any relevant contaminating source in the context of GW follow-up, to be fainter in the template image than in the search images. Therefore, we only select sources that present a difference flux that is strictly positive or within $2\sigma$ of zero across all epochs. This selection criterion leads to a final sample size of 929 sources\footnote{To check for bias from our initial selection, we also identified sources that exhibit strictly negative difference fluxes finding a comparable number (1107 sources).}.
\end{enumerate}

We note that at the time these data were taken CCD \#61 (N30) was not functioning. Additionally, during processing of the candidate list we observed severe data quality issues with CCDs \#16 (S17) and \#44 (N13) that produced a number of image artifacts and erroneous detections several orders of magnitude larger than in the other CCDs. Consequently, these three CCDs have been excluded from the analysis. This results in a $\approx 5\%$ loss of sky coverage as discussed in \autoref{tab:search}.

\begin{figure}[!t]
\begin{center}
\hspace*{-0.1in} 
\scalebox{1.}
{\includegraphics[width=0.4\textwidth]{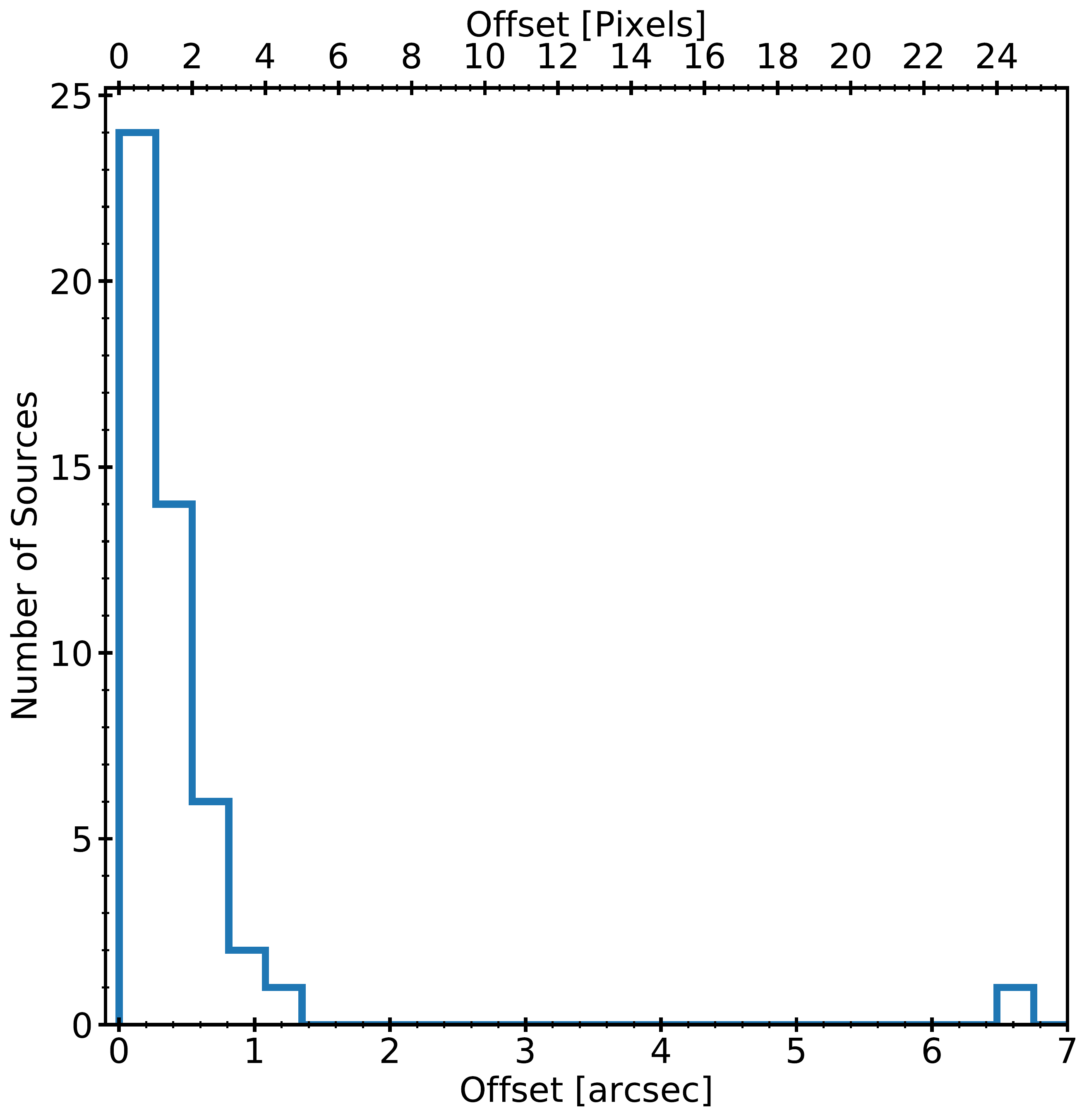}}
\caption{Offset distribution, with one pixel bins, for sources in Group 2 after visual rejection of image subtraction artifacts. Half (24 of 48) of the sources exhibit offsets smaller than 1 pixel ($0.27\arcsec$), indicating a likely AGN variability origin.}
\label{fig:offset_hist}
\end{center}
\end{figure}

\subsection{Selection Based On Template Counterpart}
\label{sec:source_cuts}

We now discuss selection criteria designed to eliminate sources that do not exhibit temporal evolution and colors expected for kilonovae. Specifically, we first leverage the fact that kilonovae are expected to have much shorter durations than most other contaminating sources (CB15). For example, a kilonova following the model of \citealt{barnes13} will fade by several magnitudes over the course of a few days, independent of the precise choice of ejecta parameters. Given that the separation in time between our last search epoch and our template epoch is 17 days, we therefore expect any kilonova-like source detected in the search images to have faded well below our detection limit in the template epoch. 

Following this reasoning, we separate our population of 929 candidates into four sub-groups based on the presence and morphology (i.e., point versus extended source) of a counterpart source in the template images. We identify such counterparts (or lack thereof) by matching the detection coordinates in our difference images against those sources detected in the {\tt SExtractor} catalogs (\autoref{sec:obs}). We define four sub-groups of the 929 sources in the following manner:

\begin{enumerate}
\item The candidate has no counterpart detection in the template epoch within a matching radius of $3\arcsec$. 

\item The candidate has an extended source counterpart in the template epoch within a matching radius of $3 \arcsec$. 

\item The candidate has a point source counterpart in the template epoch within a matching radius of $1\arcsec$, as well as an extended source counterpart brighter than 17.5 mag within a matching radius of $30\arcsec$. 

\item The candidate has a point source counterpart in the template epoch within a matching radius of $1\arcsec$, without an extended source counterpart brighter than 17.5 mag  within a matching radius of $30\arcsec$. 
\end{enumerate}

The choice of a $3\arcsec$ matching radius for Groups 1 and 2 is arbitrary, but does not ultimately affect the identification of candidates; changing the matching radius will simply shift candidates between Group 1 and Group 2. 

In Groups 3 and 4, the choice of galaxy brightness and matching radius are motivated by observations of the host galaxies of Short GRBs, which have luminosities of $\gtrsim 0.1$ L$^\star$ \citep{fong13,fong+13,berger14}. At the ALV design-sensitivity detection range $(\sim 200$ Mpc) such galaxies will be brighter than 17.5 mag. We then select the matching radius such that the probability of the point source and galaxy being associated by chance, the ``probability of chance coincidence", is $P_{\rm cc}\lesssim 0.1$. This is given by:

\begin{equation}
P_{\rm cc}(<R) = 1 - \exp{[-\pi R^2 \sigma(<m)]},
\end{equation}

\noindent where

\begin{equation}
\sigma(<m) = \left ( \frac{1}{0.33 \ln(10)} \right ) \times 10^{0.33(m - 24) - 2.44} \; {\rm arcsec}^{-2},
\end{equation}

\noindent is the expected number density of galaxies brighter than magnitude $m$ as determined by deep optical surveys (\citealt{berger10}, see also \citealt{hogg97,bloom02,beckwith06}). Therefore, setting $m = 17.5$ mag and $P_{\rm cc}(<R) = 0.1$, we find the appropriate matching radius of $R = 30\arcsec$. This radius, corresponding to a physical scale of $\approx 27$~kpc at a distance of $\approx200$~Mpc, also corresponds to some of the larger offsets measured for SGRB \citep{fong+10,fong13,berger14}. 

\subsubsection{Group 1 -- No Counterpart}
\label{sec:group1}

The first group is designed to identify sources that were detected during our search epochs but have faded below our detection threshold by the time of the template observations. Here, the matching radius $(3 \arcsec)$ is chosen such that this group of candidates will serve as the complement to the second group, as discussed above. We find that 110 of our original 929 sources belong to Group 1, but 63 of these 110 sources exhibit a $\gtrsim 3 \sigma$ detection in forced aperture photometry in the template images indicating the presence of artifacts (e.g., diffraction spikes caused by saturated stars), that are unidentified by {\tt SExtractor}. This reduces the number of candidates to 47. Manual inspection reveals these to also be image defects, specifically bad CCD columns and fringing artifacts. We therefore find no genuine transient sources that have faded beyond the limits of our template images and which have no counterparts within $3\arcsec$.

\begin{figure}[!t]
\begin{center}
\hspace*{-0.1in} 
\scalebox{1.}
{\includegraphics[width=0.36\textwidth]{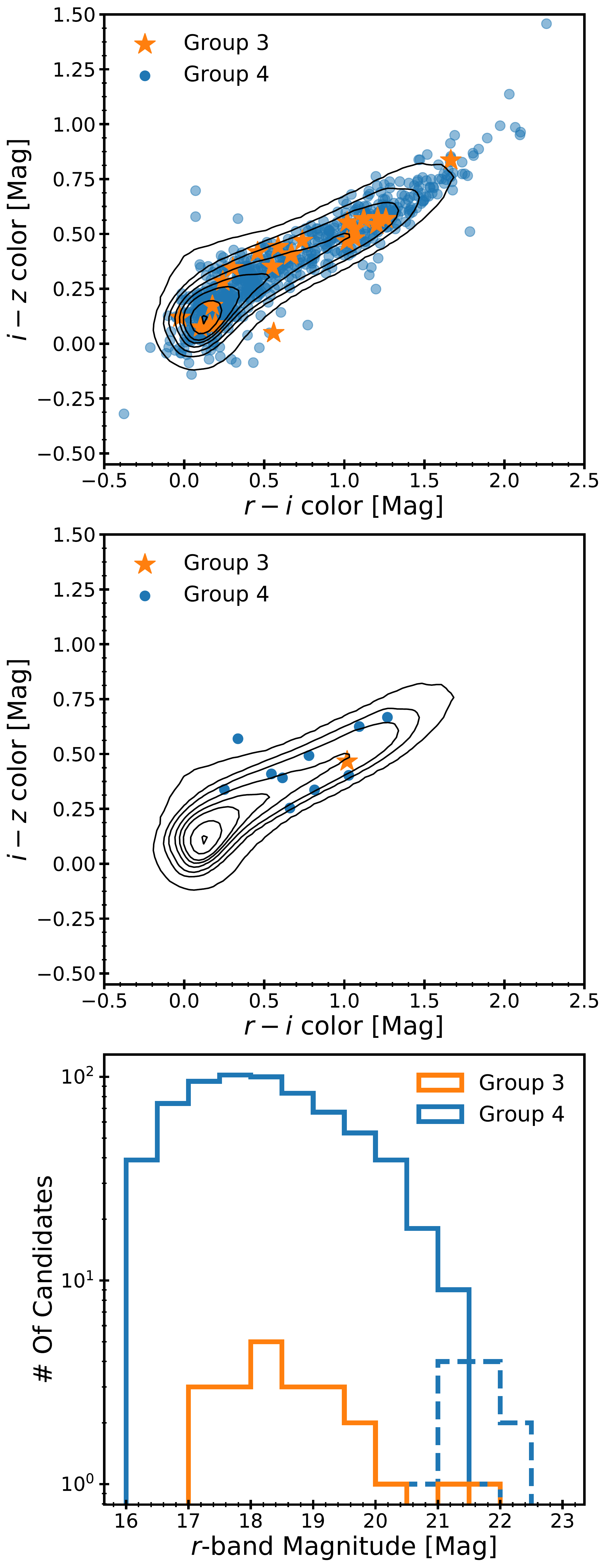}}
\caption{{\it Top:} $r-i$ vs. $i-z$ color-color diagram for the template counterparts identified in Groups 3 (orange stars; \autoref{sec:group3}) and 4 (blue circles; \autoref{sec:group4}). The black contours indicate the stellar locus in our images. We find that the majority of sources are consistent with the stellar locus indicating a variable star origin.
{\it Middle:} Same as top, but only showing those 14 sources that do not have a match in the Gaia DR1 catalog. The majority of sources are still consistent with the stellar locus, indicating that they are stellar but simply too faint to appear in the Gaia catalog.
{\it Bottom:} Distribution of $r$-band magnitudes for sources in Groups 3 and 4. The solid lines indicate sources that have a counterpart in the Gaia DR1 catalog, while dashed lines indicate sources that do not have a catalog match.}
\label{fig:color_color}
\end{center}
\end{figure}

\subsubsection{Group 2 -- Extended Source Counterpart}
\label{sec:group2}
We next search for transients that appear near to a galaxy, but like those in Group 1, have faded below the detection limit of our template epoch. We identify extended sources as those detected by {\tt SExtractor} with a {\tt CLASS\_STAR} (i.e., stellarity) value of $\le 0.8$. Applying this cut, we find that this subsample contains 94 of the original 929 sources. Visual inspection reveals that 48 of these sources are genuine transients (which we consider as potential kilonova contaminants), while the remaining 46 sources result from image subtraction artifacts in the cores of bright galaxies; we do not consider these sources in the analysis. The distribution of offsets between the transients and galaxies for the sample of 48 genuine sources is shown in \autoref{fig:offset_hist}. We find that that half (24 of 48) of the sources have an offset of $\lesssim 0.27\arcsec$ (i.e., one DECam pixel), hinting that they are nuclear in origin, and hence likely represent AGN variability.

We note that the single source in \autoref{fig:offset_hist} with an offset of $\approx6.6\arcsec$, was originally identified with an offset of $0.14\arcsec$. However, manual inspection revealed that the transient had not completely faded away in the template epoch causing {\tt SExtractor} deblending to detect a low stellarity point source still present on top of the galaxy light distribution. We correct this offset manually, but leave this source in Group 2 as that is the {\it original} classification.

\begin{figure*}[!t]
\begin{center}
\hspace*{-0.1in} 
\scalebox{1.}
{\includegraphics[width=0.95\textwidth]{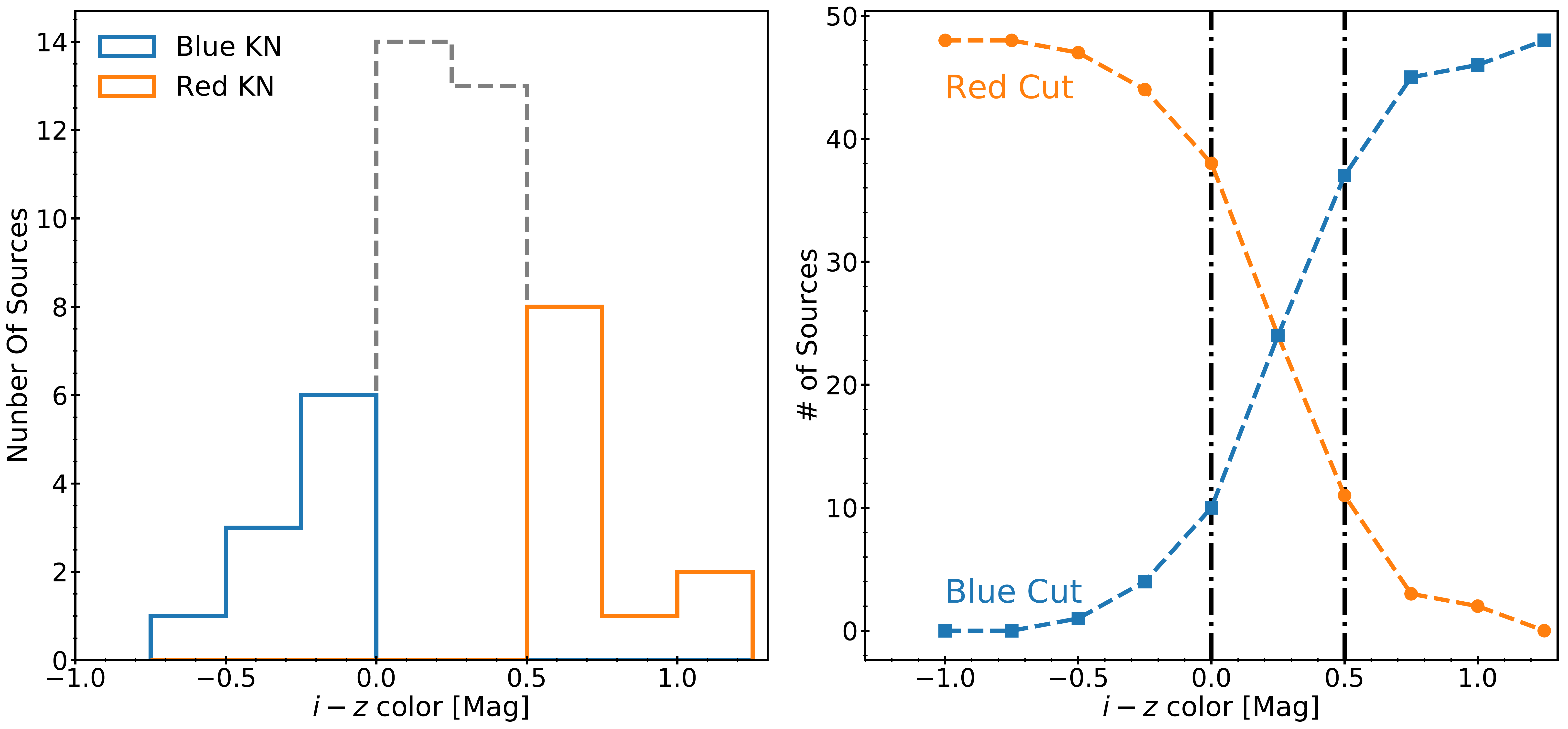}}
\caption{{\it Left:} $i-z$ color distribution for the 48 sources considered in this analysis. The 11 sources identified as red kilonova contaminants are plotted in orange (\autoref{sec:kn_search}), while the 10 sources identified as blue kilonova contaminants are shown in blue (\autoref{sec:kn_search}). The remaining 27 sources from Group 2 are plotted in grey dashed lines.
{\it Right:} The number of sources recovered by performing an $i-z$ color cut on the sample of 48 sources in Group 2. The orange line gives the number of sources redder than a given $i-z$ color. The blue line indicates the number of sources bluer than a given color cut. The vertical black lines indicate our nominal cuts of $i-z < 0$~mag and $i-z > 0.5$~mag, for blue and red kilonovae, respectively. The sharp increase in the number of sources as the chosen color threshold is relaxed can be clearly seen.}
\label{fig:color_dist_final}
\end{center}
\end{figure*}

\subsubsection{Groups 3 and 4 -- Point Source Counterparts}
\label{sec:pointsources}
We are also interested in those candidates that have a coincident point source counterpart in the template images. The presence of such a counterpart disqualifies the source as a kilonova contaminant, but this determination relies on the existence of late-time (or pre-existing) templates.  Therefore, it is still meaningful to construct a census of these sources to gain a better understanding of potential contamination in real-time GW follow-up observations, especially if pre-existing template images are not available. These sources are shown on a color-color diagram in \autoref{fig:color_color}. 

To help assess if the point source counterpart in the template images is a long-lived transient or simply a stellar variable source, we perform catalog matching for these sources against the Two Micron All-Sky Survey Point Source Catalog (2MASS PSC, \citealt{2MASS}), the Wide-field Infrared Survey Explorer all-sky release (WISE, \citealt{WISE}), and the Gaia DR1 Stellar Catalog \citep{gaia1,gaia2}\footnote{We note that Gaia DR1 uses data obtained after our data were obtained. However the time separation between the Gaia mission and our data ($> 1$ yr) along with the comparatively shallow Gaia catalogs (G $\sim$ 20 mag) makes it unlikely that any transient detected in our data will still be present in Gaia DR1.}. This matching is done using the detection coordinates from {\tt photpipe} with a matching radius of $1\arcsec$. This matching identifies 711 of the 725 candidates in Groups 3 and 4. In the following subsections we investigate the two groups separately.

\subsubsection{Group 3 -- Point Source Counterpart\\ With a Bright Nearby Galaxy}
\label{sec:group3}

We identify 23 sources that are located within $30''$ of a galaxy with $i\lesssim 17.5$ mag. \autoref{fig:color_color} shows that sources from Group 3 predominantly coincide with the stellar locus indicating a variable star origin, with the galaxy association occurring purely by chance. Given our choice of a chance coincidence probability of $P_{\rm cc} < 0.1$, it is not surprising to find 23 out of 725 spurious matches to galaxies with $i \lesssim 17.5$ mag within $30\arcsec$.

We match these 23 sources against the external catalogs (2MASS, WISE, and Gaia DR1), and find that only one source lacks a catalog match. This source is shown on the color-color diagram in \autoref{fig:color_color}, and is found to be consistent with the stellar locus. In \autoref{fig:color_color} we show the distribution of $i$-band magnitudes for Group 3 and we find that the single source not detected in the external catalogs is located at the faint end of our distribution ($r \approx 21.8$~mag). Therefore, this source is simply too faint to appear in the external catalogs.

Based on the presence of a point source counterpart in the template, association with the stellar locus, and matches to external catalogs, we conclude that none of the 23 sources in Group 3 can be considered as kilonova contaminants provided that deep template images are available. If templates are not available, then the single source without a counterpart in the Gaia DR1 catalog would be considered a kilonova contaminant.

\begin{figure*}[!t]
\begin{center}
\hspace*{-0.1in} 
\scalebox{1.}
{\includegraphics[width=0.95 \textwidth]{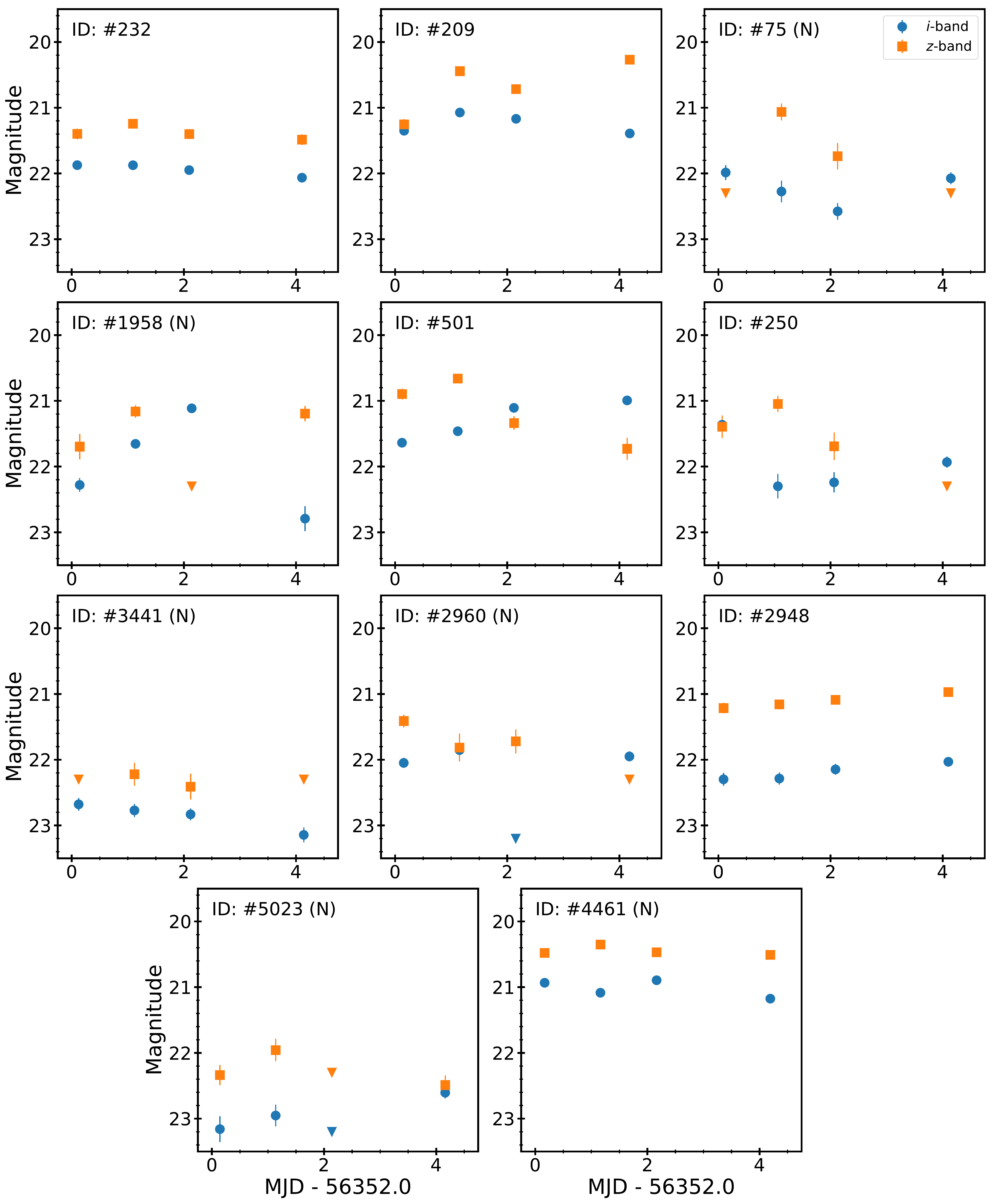}}
\caption{Light curves for the 11 sources in our red kilonova contaminant sample, constructed from the ``forced" {\tt DoPhot} PSF photometry (blue circles: $i$-band; orange squares: $z$-band). Nuclear sources are indicated by an (N) in the ID number. The $5\sigma$ limits for non-detections are indicated by triangles.}
\label{fig:final_lc_red}
\end{center}
\vspace{0.5cm}
\end{figure*}

\subsubsection{Group 4 -- Point Source Counterpart\\ Without a Bright Nearby Galaxy}
\label{sec:group4}

The remaining 702 sources do not have an associated galaxy that meets our matching criteria. We show these sources in the color-color diagram (\autoref{fig:color_color}). Similar to the candidates in Group 3, these sources coincide almost entirely with the stellar locus. The sources that extend to redder colors ($i-z \gtrsim 0.75$ mag and $r-i \gtrsim 1.75$ mag) are also consistent with the stellar locus; the contours simply do not capture this sparser region of the locus. We manually inspect sources that appear to lie outside of the stellar locus. We find that half of these sources have one or more masked/saturated pixels that bias their photometry. The remaining sources are blended or positioned in the halo of a saturated star, which affects their photometry. Finally, there is a single source with no obvious photometric issues. As we show below, this source is not detected in any external catalog, but this is likely due to its faintness ($r \approx 21.7$ mag and $g \gtrsim 23.0$ mag). Furthermore, this source does not appear near an obvious host galaxy. It is therefore most likely a faint variable star or quasar. 

Matching the sources in Group 4 against the external catalogs, we find that all but 13 of these sources have a match in 2MASS, WISE, or Gaia DR1. We show 10 of these 13 unmatched sources on the color-color diagram in \autoref{fig:color_color} and we find that they are consistent with the stellar locus.\footnote{The remaining 3 sources are not plotted due to a lack of $r$- or $z$-band detections.} Inspecting the magnitude distribution in \autoref{fig:color_color}, we find that these sources are located at the faint end of our distribution. For example, all of the sources have $r > 20.5$ mag, indicating that they are too faint to be present in the Gaia DR1 catalog. As with Group 3, we conclude that none of these sources can be considered kilonova contaminants, {\it as long as} sufficiently deep template images are available. 

\subsubsection{Summary of Initial Search}
\label{sec:search_summary}
We find 48 sources that can be considered as potential kilonova contaminants. All of these sources are from Group 2, appearing in coincidence with galaxies, and with half (24 sources) having a separation of $\lesssim 1$ pixel from the galaxy nucleus, suggesting that they are likely due to AGN variability. While it is possible for kilonovae to appear in the nuclear regions of their hosts, we generally expect larger offsets based on the fact that only $\approx 15\%$ of SGRBs exhibit offsets of $\lesssim 0.5 R_{1/2}$ \citep{fong13,fong+13,berger14}. In this analysis we do not eliminate nuclear sources from consideration, but we note that in an actual follow-up search such sources could in principle be deprioritized.

We note that rejection of the 14 sources with point-source counterparts possessing stellar colors in the template images but lacking a catalog match in Groups 3 and 4 as potential kilonova candidates is made under the assumption of template images being available. However, in the context of real-time detection in the absence of pre-existing templates such sources may lead to an additional source of contamination. We do not include these sources in our contaminant sample, because they can ultimately be rejected, but we caution that the rate of contaminants may be up to $\approx 25\%$ higher if such sources cannot be efficiently rejected.

\begin{figure*}[!t]
\begin{center}
\hspace*{-0.1in} 
\scalebox{1.}
{\includegraphics[width=0.95 \textwidth]{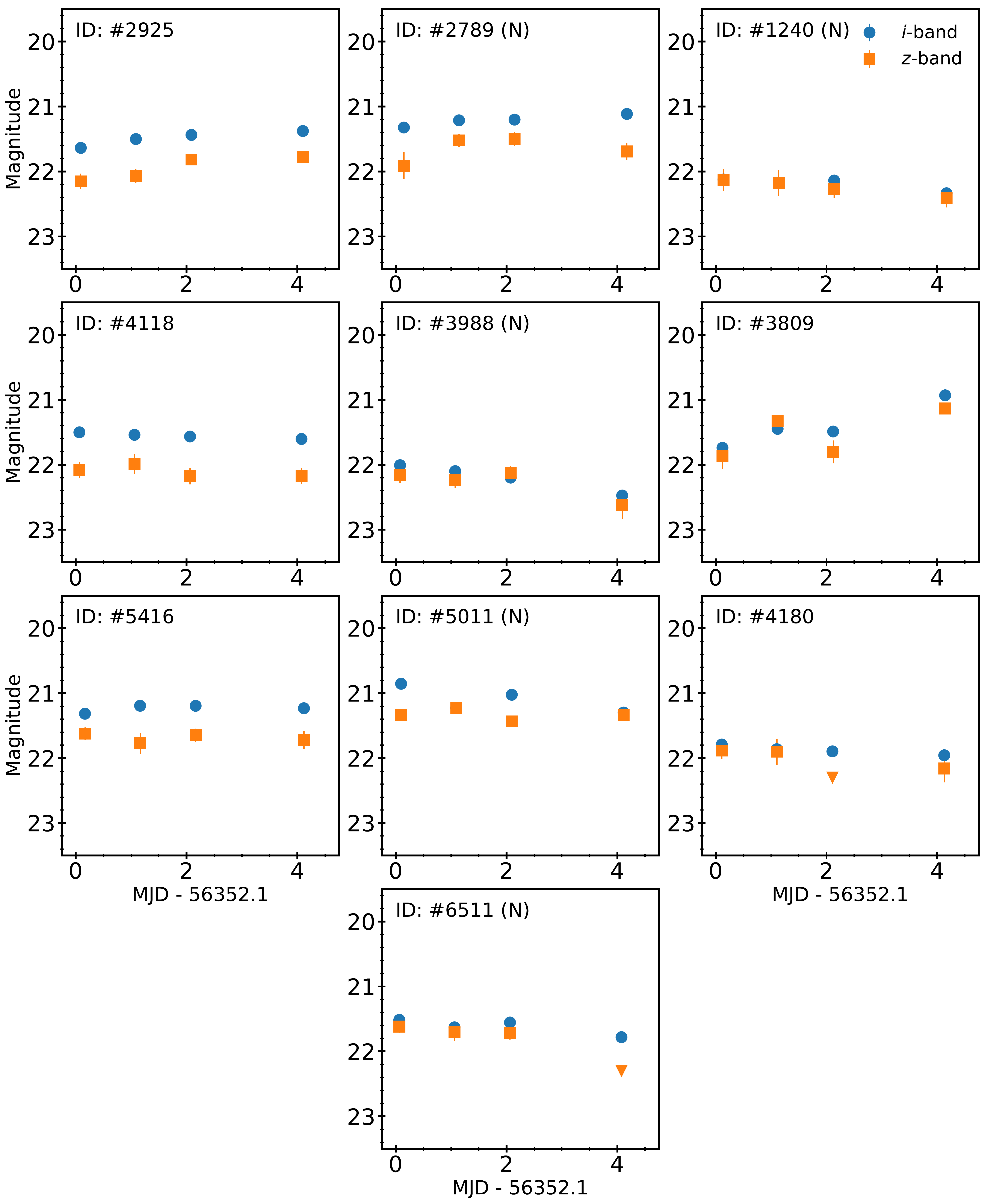}} 
\caption{Same as \autoref{fig:final_lc_red}, but for the sources in the blue kilonova contaminant sample.}
\label{fig:final_lc_blue}
\end{center}
\vspace{1.1cm}
\end{figure*}

\subsection{Color Selection of Kilonova Contaminants}
\label{sec:kn_search}
We now study the $i-z$ colors of the 48 potential contaminants to identify sources that can mimic either red or blue kilonova emission. The color distribution for all 48 sources is shown in \autoref{fig:color_dist_final}. The $i-z$ color is computed as a signal-to-noise-weighted average using the ``forced" {\tt DoPhot} photometry from the difference images. We define red kilonova contaminants as those having $i-z \gtrsim 0.5$~mag \citep[CB15]{barnes13}. For blue kilonova contaminants we require $i-z \lesssim 0.0$~mag, motivated by the models of \cite{kasen+15} for a BNS merger that results in a HMNS with $t_{\rm HMNS}\gtrsim 100$ ms. These criteria are shown in \autoref{fig:color_dist_final}. These criteria capture the tails of the color distribution, with over half of the sources (27 of 48) exhibiting $i-z = 0.0-0.5$~mag.

We find 11 sources that satisfy the color criterion for a red kilonova, of which 6 ($54\%$) are located within a pixel of a galaxy nucleus.  The light curves of all 11 sources are shown in \autoref{fig:final_lc_red}. The majority of these sources (8 of 11) have $i-z \approx 0.5-0.8$~mag and only two sources (\#75 and \#2948) have $i-z \gtrsim 1$~mag. Key aspects of the temporal evolution of red kilonova models are the rapid rise to peak ($\sim \rm few$ days) and the rapid decline post-peak ($\delta m_i \gtrsim 0.3$~mag~day$^{-1}$), in both $i$- and $z$-bands. Manually inspecting the light curves in \autoref{fig:final_lc_red}, we find no sources that clearly satisfy either of these criteria.

We find 10 sources that have colors expected for blue kilonovae. About half of these sources are located within a pixel of a galaxy nucleus. We show the light curves of all 10 sources in \autoref{fig:final_lc_blue}. The temporal evolution of blue kilonova models is more rapid, with a shorter duration, than that of red kilonovae. We do not find any systematic trends in light curve behavior for the 10 sources, or when inspecting the nuclear and non-nuclear sources separately.

The complete set of selections are summarized in \autoref{tab:cuts_sum}. We selected subsets of red and blue kilonova contaminants with specific color cuts, but we note that the models motivating these choices have uncertainties that can affect the kilonova colors (e.g., ejecta mass and velocity, ejecta composition, uncertainties in $r$-process opacities, etc.). This makes understanding the effect of color criteria on the size of the contaminant sample crucial. 

The number of red and blue kilonova contaminants as a function of color is shown in \autoref{fig:color_dist_final}. We find that the number of sources in either sample increases significantly if the selection on $i-z$ color is relaxed. For example, if we search for red kilonova contaminants by requiring $i-z \gtrsim 0.3$~mag, the number of contaminants rises to 22, a twofold increase. Similarly, if we relax our color selection for blue kilonova to $i-z \lesssim 0.2$~mag, the number of contaminants rises to 20, again a twofold increase over the original sample. 

\subsection{A Search for Contamination on Nightly Timescales}
\label{sec:kn_short}
We also search our data for sources that could appear as contaminants on the timescales relevant to the short-lived ``neutron precursor," that are speculated to accompany some mergers (\autoref{sec:kn_models}, \citealt{metzger15}). We accomplish this by leveraging the rapid cadence of our observations to identify potential contaminants that are detected during a single night, or just a half-night epoch, probing transient and variable events that occur with timescales of $3$ hr to 1 day and $\lesssim3$ hours, respectively. We search for candidates based on their behavior in the ``forced" {\tt DoPhot} photometry as follows:

\begin{enumerate}
\item We search for transients with a characteristic timescale of 3 hours to 1 day by selecting candidates that exhibit two $\gtrsim 10\sigma$ detections in $i$-band during a single night of observations (i.e., two epochs). Outside of these epochs, the sources must exhibit an $i$-band difference flux that is a factor of $\gtrsim 10$ fainter than the maximum $i$-band difference flux measured during the night of interest. This requirement is consistent with the rapid fading expected for ``neutron precursors" \citep{metzger15}.

\item We search for transients with characteristic timescales of $\lesssim 3$ hours by selecting candidates that exhibit a single $\gtrsim 10\sigma$ detection in both $i$- and $z$-band in a single epoch. Outside of this epoch, the sources must exhibit a difference flux that is a factor of $\gtrsim 10$ fainter than the difference flux measured during the epoch of interest, in both $i$- and $z$-bands, again motivated by the rapid fading expected for ``neutron precursors" \citep{metzger15}.
\end{enumerate}

We find 9 sources with a timescale between 3 hours and 1 day. We perform a manual inspection of this sample, finding 5 genuine sources and 4 that result from image subtraction artifacts. Matching the 5 sources to our template images and external catalogs, we find that all of them have a point-source match in both our template images and Gaia DR1, and hence represent stellar variability or flaring. We find no evidence for extragalactic contamination from sources with a timescale between 3 hours and 1 day.

We find 39 sources that match our selection criteria for a duration of $\lesssim 3$ hr. We manually inspect these sources and find 24 genuine sources, with the remaining candidates resulting from image subtraction artifacts. Fifteen of these sources have point source counterparts in our template images, as well as matches in the Gaia DR1 catalog. An additional six sources are matched to high stellarity sources, but these are all likely too faint ($r \gtrsim 21.5$~mag and $g \gtrsim 23.0$~mag) to be present in the Gaia DR1 catalog. The remaining 3 sources exhibit significant trailing in at least one epoch and no detected counterpart in the template images, indicating that they are asteroids. Thus, we find no non-stellar or non-moving sources with a timescale of $\lesssim 3$ hr.

\begin{figure*}[!t]
\begin{center}
\hspace*{-0.1in} 
\scalebox{1.}
{\includegraphics[width=0.95 \textwidth]{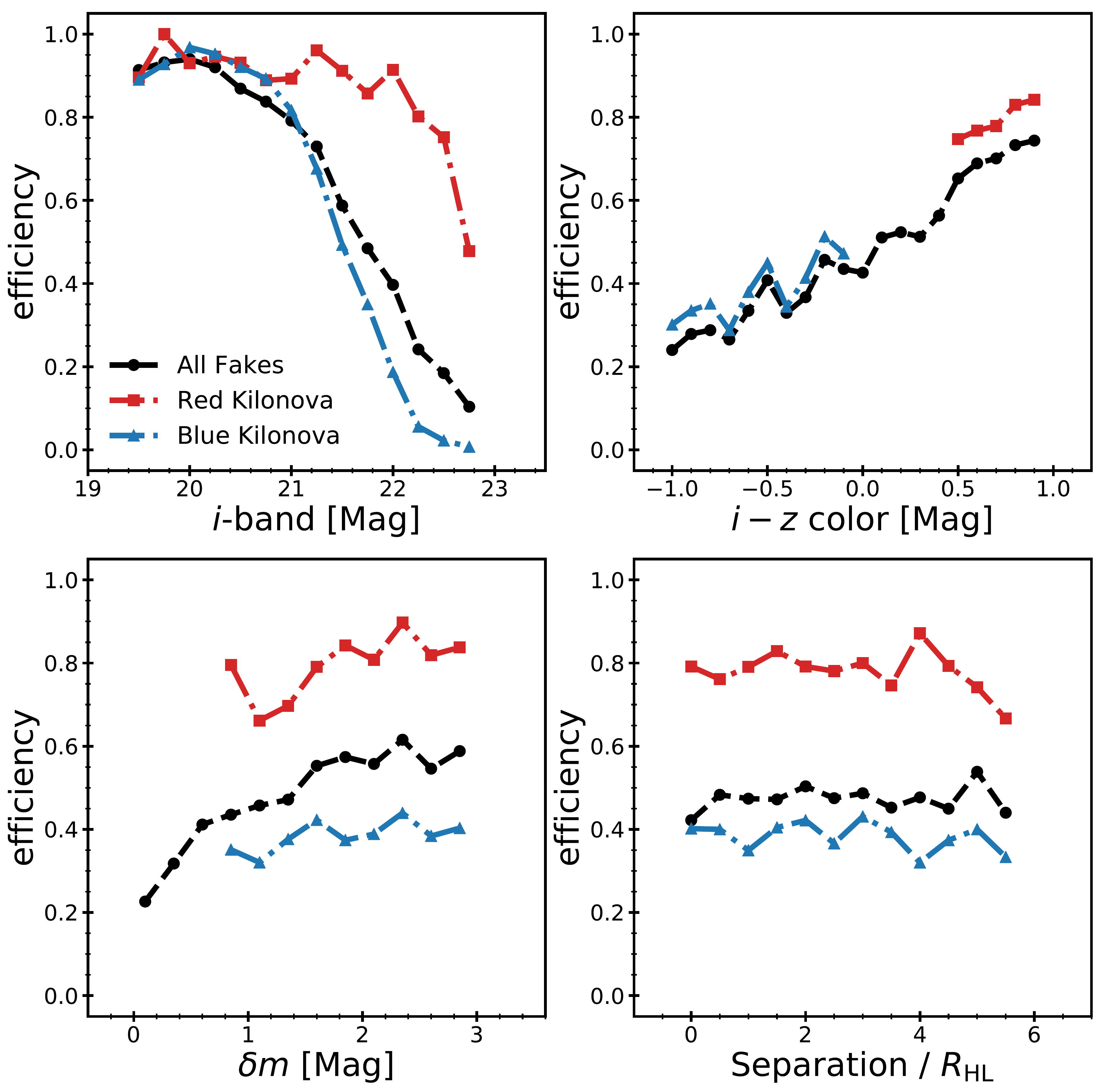}}  
\caption{Plots of recovery efficiency as a function of various fake source injection parameters. The efficiency for the entire population of fake sources is shown as a black line while the efficiency for red kilonova and blue kilonova fake sources are shown as orange and blue lines, respectively. {\it Top Left:} Efficiency as a function of $i$-band magnitude. We note that our efficiency for red kilonova fake sources is higher than our mean efficiency, while the blue kilonova efficiency declines more rapidly compared to the red kilonova fake sources. This is due to the design of our observations, which are aimed at red sources. {\it Top Right:} Efficiency as a function of $i-z$ color. We are more sensitive to objects that are red in $i-z$ color, which drives the efficiency differences seen in the other panels. {\it Bottom Left:} Efficiency as a function of fading ($\delta m$, see text for definition). {\it Bottom Right:} Efficiency as a function of host galaxy separation. We find no dependence on separation.}
\label{fig:eff_1D}
\end{center}
\end{figure*}

\section{Detection Efficiency}
\label{sec:fakes}
To determine the areal rate of the various kilonova contaminants we need to determine the detection efficiency of our search method. We accomplish this by injecting point sources into both our search and template images. We inject each source with a constant brightness and $i-z$ color in the search images. To assess the impact of residual flux in the template images we use a range of fading levels between the search and template images.  Finally, to assess the effect of host galaxy brightness on our recovery efficiency we inject the sources on and near galaxies identified using {\tt SExtractor} photometry on the template epoch. 

We inject ten point sources around 580 galaxies identified across the 21 fields in our dataset for a total of 5800 injected sources. The population is constructed as follows:

\begin{enumerate}
\item We select extended sources by requiring a half-light radius of $R_{1/2} > 20$ pixels and a stellarity value of $<0.2$. The choice of $R_{1/2}$ corresponds to the approximate size of a Milky Way like galaxy at a distance of $\approx 200$ Mpc, appropriate for NS-NS mergers detections by aLIGO. These values are determined from the {\tt FLUX\_RADIUS} and {\tt CLASS\_STAR} parameters in the {\tt SExtractor} catalog \citep{bertin96}, and verified by manual inspection.

\item We inject 10 sources at random locations around each galaxy, constrained to a box that is $4R_{1/2}$ on a side.

\item We inject sources with an $i$-band magnitude range of $19.5-23$ mag, with a volume weighting to produce a realistic distribution of faint sources.

\item We assign each source a color of $i-z=-1$ to $1$ mag, with a uniform distribution.

\item We assign each source a difference in magnitude between the science and template images of $\delta m= 0.2-3$ mag, with a uniform distribution; the range is designed to capture slow fading that would be typical of supernovae and rapid fading typical of kilonovae.

\end{enumerate}

To recover and study the injected sources we process the data in the same manner as described in \autoref{sec:obs}, and apply the data quality selection criteria as described in \autoref{sec:data_cuts}. We then match the identified sources against the list of injected sources, allowing for an astrometric match tolerance of 2 pixels.

For the purpose of determining our detection efficiency in a manner relevant to our search for kilonova contaminants we identify two primary groups of injected sources, namely those that have red kilonova properties (i.e., $i-z>=0.5$ mag and $\delta_m>=1$ mag) and those that have blue kilonova properties (i.e., $i-z<=0.0$ mag and $\delta_m>=1$ mag). These criteria lead to 1006 and 1955 injected sources, respectively. The remaining 2839 sources span a range of properties intermediate between kilonova and supernova properties. We consider the effect of the source brightness, color, fading, and the separation from the host galaxy on our ability to detect sources. We define our efficiency as the ratio of the number of sources recovered to the number of sources injected.

In \autoref{fig:eff_1D} we plot the detection efficiency as a function of $i$-band magnitude for the full sample, as well as for the subsets of red and blue kilonova sources. We find an overall efficiency of $\gtrsim0.8$ for $i \lesssim 21.5$ mag and 0.5 at $i \approx 22$ mag. For red kilonova sources we find a higher efficiency of $\approx 0.9$ at $i\lesssim 22$ mag, and 0.5 at $i\approx 22.8$ mag. Our efficiency for blue kilonova sources is $\approx 0.9$ for sources with $i\lesssim 21$ mag, and 0.5 at $i\approx 21.5$ mag. The higher efficiency for red kilonova sources is due to the relative depths of our $i$- and $z$-band images, which were chosen to explore red sources.

\begin{figure}[!t]
\begin{center}
\hspace*{-0.1in} 
\scalebox{1.}
{\includegraphics[width=0.45 \textwidth]{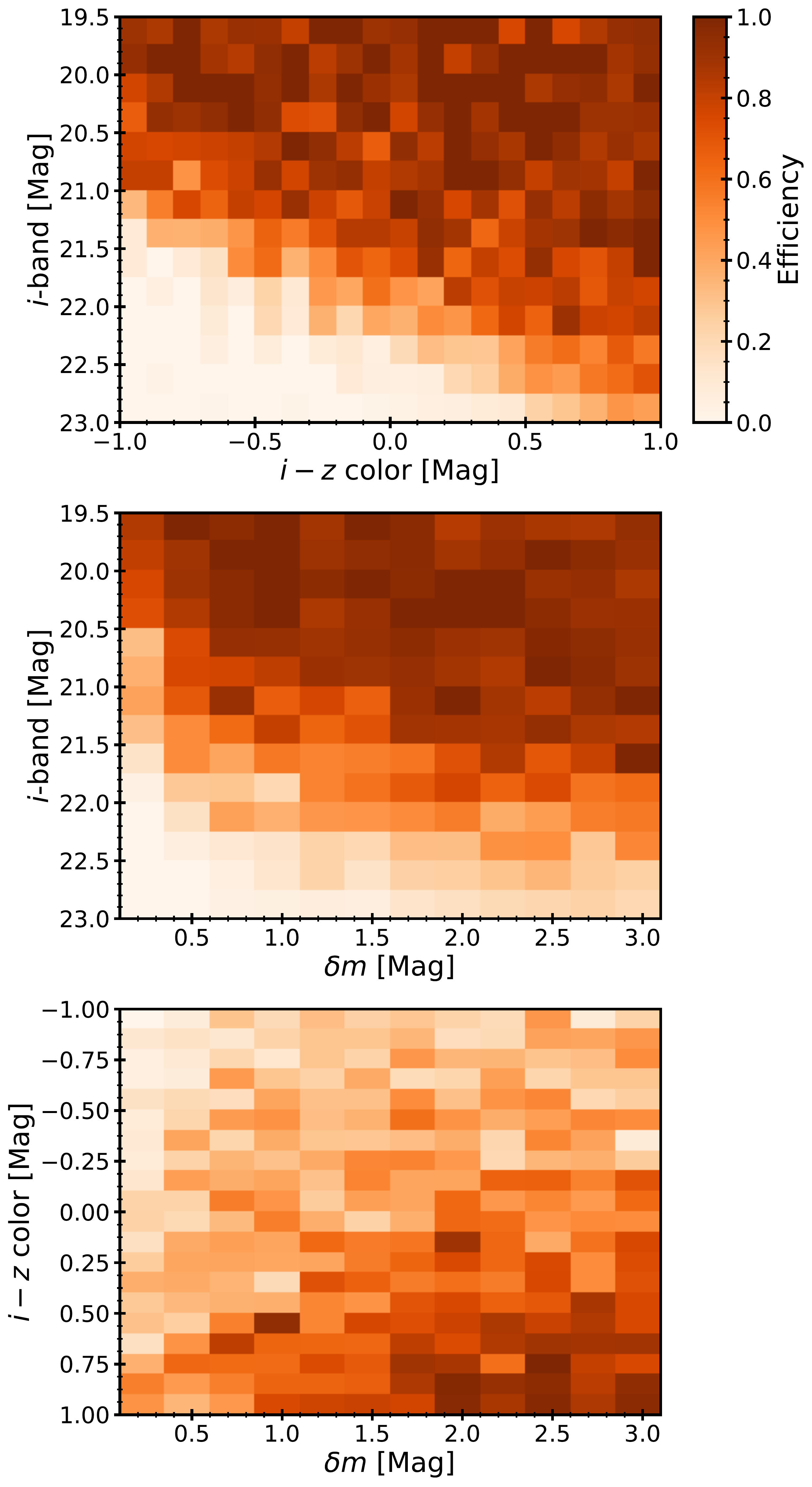}}
\caption{Two dimensional histograms of efficiency. {\it Top:} Efficiency as a function of $i$-band magnitude and $i-z$ color. There is a clear dependence of depth on color, due to the design of our observations. {\it Middle:} Efficiency as a function of $i$-band magnitude and source fading. There is a sharp decline in efficiency for injected sources with a fading of $\lesssim 1$ mag between the search and template images, but otherwise our efficiency is constant above this value. {\it Bottom:} Efficiency as a function of $i-z$ color and source contrast. Our highest efficiency is for sources with red $i-z$ colors and large contrast, the properties expected for red kilonovae.}
\label{fig:eff_2D}
\end{center}
\end{figure}

The efficiency as a function of color for all magnitudes and fading rates is also shown in \autoref{fig:eff_1D}. We find that the efficiency is $\lesssim 0.5$ for $i-z<0$ mag, and then increases monotonically to $\approx 0.8$ by $i-z\approx 1$ mag. For red kilonova sources the efficiency is $\approx10\%$ higher than for the general population of injected sources, while for blue kilonova sources it is comparable to that for the full sample.

We next explore the efficiency as a function of fading ($\delta m$) across all colors and magnitudes. We find that the efficiency is $\lesssim 0.5$ for mild fading of $\delta m\lesssim 0.5$ mag, but then steadily increases to about 0.7 when $\delta m\approx 3$ mag. For red kilonova sources, the efficiency is about 20\% higher than for the full sample, while for blue kilonova sources it is approximately $15$\% lower than for the general population of injected sources.

Finally, we investigate the efficiency as a function of angular separation, normalized by $R_{1/2}$, between the injected source and the galaxy. We find that the efficiency is relatively constant for the full range of separations, spanning $\approx 0-5 R_{1/2}$. This indicates that our recovery efficiency is uniform even at negligible separations from galaxy centers. For red kilonova sources, the efficiency is about 30\% higher than for the total sample of injected sources, while for blue kilonova sources it is about 10\% lower than for the full sample.

In general, the final efficiency for a given source population depends on a combination of source properties. Histograms of the two-dimensional efficiency as a function of multiple source properties are shown in \autoref{fig:eff_2D}.

We quantify our final efficiencies at a limiting magnitude of $i \lesssim 22.5$ mag, corresponding to the magnitude at which the efficiency for the entire population of fake sources is $\lesssim0.2$. We compute final efficiencies of $\approx90\%$ for red kilonova sources and $\approx60\%$ for blue kilonovae sources.

The initial set of cuts to determine if a fake point source is red or blue kilonova-like are applied to the injected properties, but the measured properties can vary quite significantly. For faint sources ($i \gtrsim 22.5$ mag), the measured color of a source can be inaccurate by $\gtrsim 0.25$~mag and the error does not approach zero until $i  \lesssim 20$ mag. This error in color can lead to a fake source being miscategorized during recovery. We investigate this effect by identifying sources in our sample that would be miscategorized if our analysis was based on the {\it measured} source properties. We find that this is an overall minor effect, leading to a $\lesssim 10 \%$ change to the calculated efficiency.

\section{Contamination Rates}
\label{sec:contam_rates}
\begin{figure}[!t]
\begin{center}
\hspace*{-0.1in} 
\scalebox{1.}
{\includegraphics[width=0.40 \textwidth]{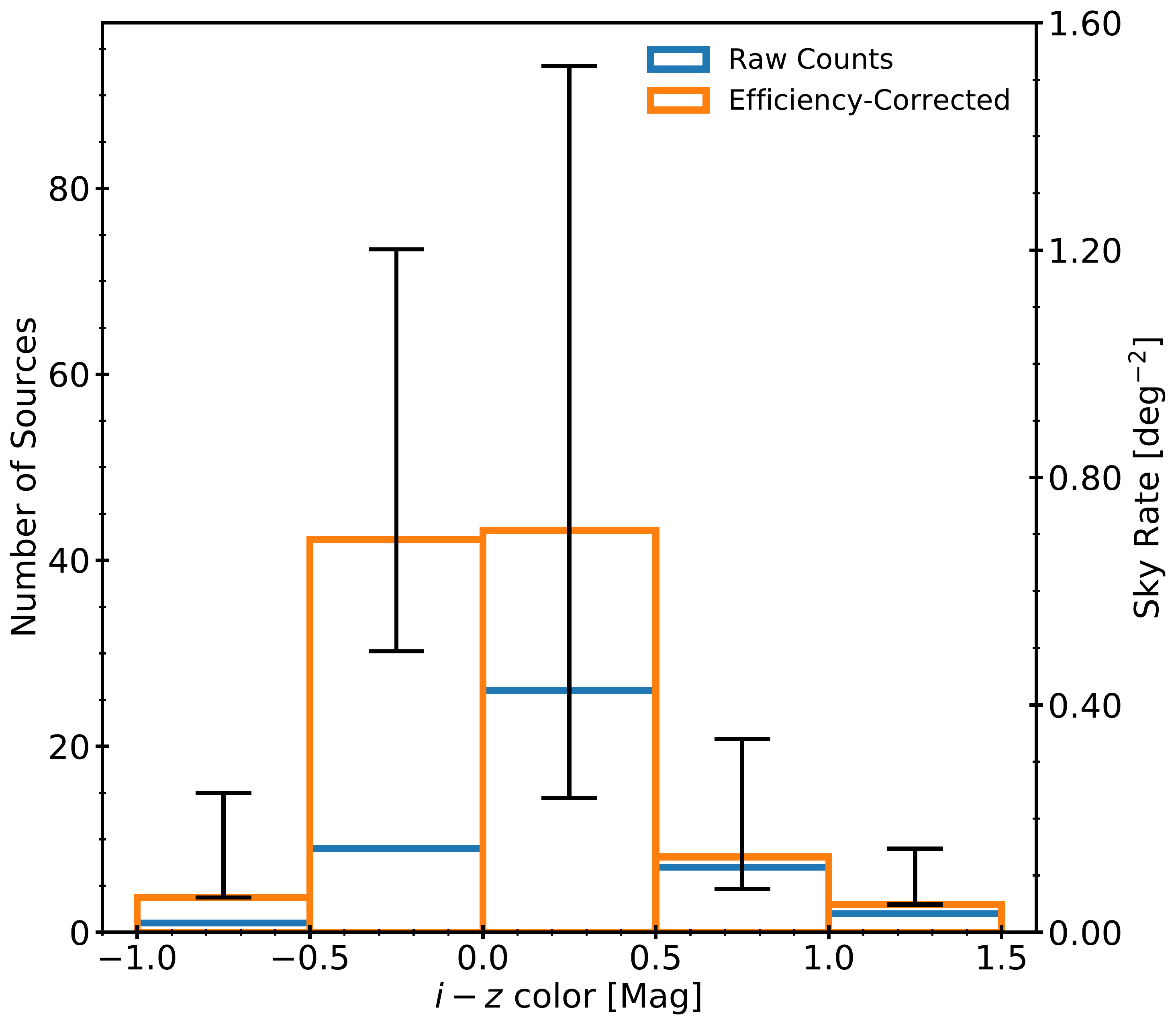}}
\caption{Histogram of contaminant numbers and areal rates as a function of $i-z$ color. The blue lines indicate the raw source counts, while the orange line indicates the efficiency-corrected counts. The error bars represent the $1\sigma$ confidence interval as computed assuming Poisson counting statistics.}
\label{fig:color_dist_corr}
\end{center}
\end{figure}

We now combine the results of Sections~\ref{sec:search} and~\ref{sec:fakes} to compute the areal rate of contaminating sources for the effective sky area for our search ($A_{\rm sky} \approx56$~deg$^2$). We note that our rates are for the relevant ``per search," and not per unit time. These rates can easily be combined with the size of a given GW localization region to compute an expected number of contaminating sources. We first compute the expected detection efficiency relevant for each source in our sample given its color and magnitude; we do not consider the source location relative to a galaxy since our efficiency is uniform with angular separation (\autoref{fig:eff_1D}). In \autoref{fig:color_dist_corr} we plot the efficiency-corrected number of sources as a function of $i-z$ color in three bins corresponding to red kilonovae, blue kilonovae, and intermediate colors. We compute $1\sigma$ confidence intervals assuming simple Poisson counting statistics. These confidence intervals are computed for the raw number of sources and are then scaled by the mean efficiency in each color bin.

We maintain consistency with our calculated efficiencies from \autoref{sec:fakes} by only considering sources with a mean $i$-band magnitude of $i \lesssim 22.5$~mag as computed from the forced {\tt DoPhot} photometry. At this magnitude limit there are $45$ sources out of the total initial sample of $48$ from \autoref{sec:kn_search}. The magnitude-limited sample of red kilonova contaminants found in \autoref{sec:kn_search} is composed of 9 sources. Correcting for detection efficiency leads to a contaminant rate of 0.16 deg$^{-2}$ at a $5\sigma$ limiting magnitude of $i \lesssim 22.5$ mag. The efficiency-corrected rate of non-nuclear red kilonova contaminants is 0.11 deg$^{-2}$.  For a typical ALV localization region of $\sim 100$ deg$^2$ we therefore expect 16 kilonova contaminants, with about 11 being non-nuclear (at $i\lesssim 22.5$ mag).

For the blue kilonova contaminants all 10 sources are above the magnitude limit. The efficiency corrected number is 46, driven primarily by two sources that are faint ($i \gtrsim 22.2$ mag) and blue ($i-z \approx -0.1$~mag) and therefore have low detection efficiencies of only $\approx 0.1$. If we remove these sources from consideration, then the efficiency-corrected number is 20. There is therefore about a factor of 2 uncertainty in the resulting contamination rate. Considering all sources, the contamination rate is about 0.80 deg$^{-2}$ to a limiting magnitude of $i\approx 22.5$ mag, or $\approx 80$ sources in a $100$ deg$^2$ localization region. The complete set of selections and efficiency-corrected rates are presented in \autoref{tab:cuts_sum}.

\begin{deluxetable}{l c c c c}
\tabletypesize{\footnotesize}
\tablecolumns{5}
\tablewidth{0pt}
\tablecaption{Summary of Final Contaminant Sample at $i\lesssim 22.5$ mag
\label{tab:cuts_sum}}
\tablehead{
	\colhead{Selection} &
	\colhead{$N$ (Raw)} &
	\colhead{$\mathcal{R}$ (Raw)} &
	\colhead{$N$ (Corrected)} &
	\colhead{$\mathcal{R}$ (Corrected)} \\
	\colhead{} &
	\colhead{} &
	\colhead{(deg$^{-2}$)} &
	\colhead{} &
	\colhead{(deg$^{-2}$)}
}
\startdata
Total Sample & 45$_{-9}^{+10}$ & 0.80$_{-0.16}^{+0.18}$ & 101$_{-14}^{+15}$ & 1.79$_{-0.25}^{+0.26}$ \\
Nuclear & 21$_{-6}^{+ 7}$ & 0.38$_{-0.11}^{+0.12}$ & 58$_{-10}^{+12}$ & 1.03$_{-0.19}^{+0.20}$ \\
Non-Nuclear & 24$_{-7}^{+ 7}$ & 0.43$_{-0.12}^{+0.13}$ & 43$_{-8}^{+10}$ & 0.76$_{-0.15}^{+0.17}$ \\
Red & 9$_{-4}^{+ 4}$ & 0.16$_{-0.07}^{+0.07}$ & 12$_{-4}^{+ 5}$ & 0.20$_{-0.07}^{+0.09}$ \\
Blue & 10$_{-4}^{+ 5}$ & 0.18$_{-0.07}^{+0.09}$ & 46$_{-8}^{+11}$ & 0.82$_{-0.16}^{+0.18}$ \\
Red/Nuclear &  4$_{-3}^{+ 3}$ & 0.07$_{-0.05}^{+0.05}$ &  5$_{-2}^{+ 4}$ & 0.09$_{-0.05}^{+0.05}$ \\
Blue/Nuclear &  5$_{-3}^{+ 3}$ & 0.09$_{-0.05}^{+0.05}$ & 34$_{-7}^{+ 9}$ & 0.60$_{-0.14}^{+0.15}$ \\
Red/Non-Nuclear &  5$_{-3}^{+ 3}$ & 0.09$_{-0.05}^{+0.05}$ &  7$_{-3}^{+ 4}$ & 0.11$_{-0.06}^{+0.07}$ \\
Blue/Non-Nuclear &  5$_{-3}^{+ 3}$ & 0.09$_{-0.05}^{+0.05}$ & 13$_{-4}^{+ 5}$ & 0.22$_{-0.08}^{+0.08}$ \\
Timescale: $3-24$ hr & $\lesssim 3$ & $\lesssim 0.05$ & $\lesssim 4$ & $\lesssim 0.07$ \\
Timescale: $\lesssim3$ hr & $\lesssim 3$ & $\lesssim 0.05$ & $\lesssim 4$ & $\lesssim 0.07$ \\
\enddata
\tablecomments{Summary of selections made for our magnitude-limited final sample of 45 sources, including $1 \sigma$ errors on source counts. We give raw and efficiency corrected number of sources $(N)$ and sky rate $(\mathcal{R})$, assuming our search represents a typical region of sky. We define red and blue sources as those with $i-z > 0.5$~mag and $i-z < 0.0$~mag, respectively. We define nuclear sources as those exhibiting an offset from the nucleus of their host galaxy of $\lesssim 1$ pixel ($\lesssim 0.27\arcsec$).}
\end{deluxetable}

\begin{deluxetable*}{c | c c c c | c c c c | c}
\tabletypesize{\footnotesize}
\tablecolumns{10}
\tablewidth{0pt}
\tablecaption{Summary of O1 Optical Follow-Up
\label{tab:O1_sum}}	
\tablehead{
    \colhead{} &
    \multicolumn{4}{|c|}{GW150914} &
    \multicolumn{4}{|c|}{GW151226} &
    \colhead{} \\
    \colhead{Group} &
    \multicolumn{1}{|c}{$m_{5\sigma}$} &
    \colhead{$A_{\rm Sky}$} &
    \colhead{$N$} &
    \multicolumn{1}{c|}{$\mathcal{R}_{\rm sky}$} &
    \multicolumn{1}{|c}{$m_{5\sigma}$} &
    \colhead{$A_{\rm Sky}$} &
    \colhead{$N$} &
    \multicolumn{1}{c|}{$\mathcal{R}_{\rm sky}$} &
    \colhead{Notes} \\
    \colhead{} &
    \multicolumn{1}{|c}{(Mag)} &
    \colhead{(deg$^2$)} &
    \colhead{(Number)} &
    \multicolumn{1}{c|}{(deg$^{-2}$)} &
    \multicolumn{1}{|c}{(Mag)} &
    \colhead{(deg$^2$)} &
    \colhead{(Number)} &
    \multicolumn{1}{c|}{(deg$^{-2}$)} &
    \colhead{}
}
\startdata
DECam & $i \lesssim 22.5$ & 102 & 9 &  0.08 & $i \lesssim 21.7$ & 28.8 & 4 & 0.13 & {\it A} \\
iPTF & $r \lesssim 20.5$ & 126 & 8 & 0.06 & $r \lesssim 20.5$ & 731 & 21 & 0.03 & {\it B} \\
J-GEM/KWFC  & $i \lesssim 18.9$ & 24 & 0 & $\lesssim0.13$ & $r \lesssim 20.5$ &  778 & 13 & 0.02 & {\it C} \\
Pan-STARRS & $i \lesssim 20.8$ & 442 & 56 & 0.12 & $ i \lesssim 20.8$ & 290 & 49 & 0.17 & {\it D} \\
\enddata
\tablecomments{Summary of optical follow-up for the two high-signficance GW events detected during
the first aLIGO observing run \citep[O1;][]{gw150914_em, gw150914_em_supp}. We report the published limiting magnitude $(m_{5\sigma})$, the area covered
($A_{\rm sky}$), the number of reported candidates $(N)$, and the projected sky rate $(\mathcal{R} \equiv N / A_{\rm sky})$.}
\tablenotetext{A }{~References: \cite{ss+16,cowp16}}
\tablenotetext{B }{~References: \cite{kasliwal+16, cenko15,cenko16}}
\tablenotetext{C }{~References: \cite{morokuma+16,yoshida+17} \\
Note: Here we only consider the KWFC wide-field survey component of the J-GEM follow-up.}
\tablenotetext{D }{~References: \cite{smartt+16a,smartt+16b}}
\end{deluxetable*}

We compare the contamination rates derived from our search to those from several follow-up observations of GW sources from the first aLIGO observing run. We focus on optical follow-up using wide-field instruments for GW150914 and
GW151226. Specifically, we use the published results from observations with DECam (\citealt{ss+16,cowp16}), the intermediate Palomar Transient Factory (\citealt{kasliwal+16,cenko15,cenko16}), the Kiso Wide-Field Camera (KWFC) used as part of the J-GEMs collaboration (\citealt{morokuma+16,yoshida+17}), and the Pan-STARRS/PESSTO/ATLAS search \citep{smartt+16a,smartt+16b}. The parameters and results of these searches are summarized in \autoref{tab:O1_sum}. 

It is important to note that the searches conducted in response to GW150914 and GW151226 were fundamentally different from our study, as well as from searches that would be required to detect actual kilonovae. These searches employed slow cadences ($\approx \rm few$ days), shallower depths, and did not use colors for source selection. Nevertheless, we can use the numbers of reported transients, which were all deemed unrelated to the GW event, as a proxy for the contamination rate.

We find that our measured contaminant rate, for both red and blue sources, is higher than those reported during O1 follow-up, which had a typical rate of  $\lesssim 0.1$ deg$^{-2}$. However direct comparison requires careful consideration of selection criteria and depth. For example, the iPTF and KWFC follow-up of GW151226 both achieved comparable depth to each other and their reported contaminant rates are in good agreement. By comparison, the DECam and iPTF follow-up of GW150914, both observed the same contamination rate despite the DECam observations being significantly deeper. This is due to a difference in selection criteria as the DECam search focused only on finding rapidly declining transients.

It is critical to note that while our measured contamination rate is higher, the observations used in this work were conducted at the depths and cadences necessary for searches targeting kilonovae. Therefore, the rates derived here are more relevant for GW follow-up conducted in response to BNS and NS-BH mergers. 

\subsection{Rates for Short Timescale Transients}
\label{sec:rates_short}

We also compute expected contamination rates for sources identified in our short timescale search (\autoref{sec:kn_short}). For both the $\approx 3-24$ hr and $\lesssim 3$ hr timescales we did not identify any credible source of extragalactic contamination. This leads to an efficiency-corrected upper limit of $\lesssim 0.07$ deg$^{-2}$ (95\% confidence level) at $i\approx 22.5$ mag for both populations. A $100$ deg$^2$ GW localization region would therefore contain $\lesssim 7$ contaminants on these timescales. These rates are included in \autoref{tab:cuts_sum}.

These rates can be compared to the Pan-STARRS1 Medium-Deep Survey fast transients search of \cite{berger+13}. That study focused on a timescale of $\sim 0.5$ hr to $\sim 1$ d and led to an upper limit on the rate of extragalactic transients of $\lesssim 2.4\times 10^{-3}$ deg$^{-2}$ day$^{-1}$ for the timescale of 1 d, and $\lesssim 0.12$ deg$^{-2}$ day$^{-1}$ for a timescale of 0.5 hr at a limiting magnitude of $r\approx 22.4$ mag. For our search (56 deg$^2$ and 4 d) we therefore expect $\lesssim 0.5$ events on a timescale of $\lesssim 1$ d, consistent with our non-detection of any extragalactic contaminants.

\section{Discussion and Conclusions}
\label{sec:conc}
We presented an empirical study of contamination rates in rapid, deep, wide-field optical follow-up of GW sources. Our observations used DECam to cover a wide search area of $56$ deg$^2$ and to probe the regime applicable for kilonovae at expected BNS merger detection distances.  We also explored timescales ranging from $\sim 3$ hours (applicable for a neutron precursor signal) to several days (applicable to blue and red kilonova emission). We search the data for transient sources that would contaminate searches for red kilonova emission, blue kilonova emission, or a neutron precursor.  We note that the former is a robust prediction of BNS mergers, while the latter two are more speculative and depend on currently unknown factors such as the neutron star equation of state. The key results of our study are as follows:

\begin{enumerate}
\item We find 48 transient sources coincident with galaxies, and lacking a point source in the template images.  Furthermore, we find 14 transients with point source counterparts that exhibit stellar colors, but are too faint to be present in catalogs such as Gaia DR1.  These sources can be rejected as contaminants under most circumstances, but may confuse real-time searches if pre-existing templates are not available.

\item We use $i-z$ color selection for the 48 sources to identify contaminants for red and blue kilonovae \citep{barnes13,kasen+15}. We find 11 red kilonova ($i-z \gtrsim 0.5$~mag) and 10 blue kilonova ($i-z < 0$~mag) contaminants.

\item We search the data for transients with a timescale of $\lesssim 1$ d, which will contaminate searches for a ``neutron precursor" signal.  We identify no credible evidence for extragalactic contamination on these timescales.

\item We compute efficiency-corrected areal rates for contaminants (per GW follow-up search) at a limiting magnitude of $i\approx 22.5$ mag, of $\mathcal{R}_{\rm tot}\approx 1.79$ deg$^{-2}$, $\mathcal{R}_{\rm red}\approx 0.16$ deg$^{-2}$, and $\mathcal{R}_{\rm blue}\approx 0.80$ deg$^{-2}$. We compute an upper limit areal rate for sources with a characteristic timescale of $\lesssim 1$ day of $\lesssim 0.07$ deg$^{-2}$ at the 95\% confidence level.

\item Our derived contamination rates are higher than reported optical transients found in follow-up of GW150914 and GW151226 ($\mathcal{R} \approx 0.1$ deg$^2$), but this is due to the greater depth of our observations, which are better matched to kilonova detections (CB15).

\end{enumerate}

For three detectors (ALV), the typical localization regions are $\lesssim 100$ deg$^2$. For ALV at design sensitivity, about half of all BNS mergers are expected to be localized to $\lesssim10$ deg$^2$ \citep{chen16}. Similarly, with a five-detector network (ALV+KAGRA+LIGO India) it is expected that about 90\% of BNS mergers will be localized to $\lesssim 10$ deg$^2$ \citep{chen16}. Based on our derived rates, we expect such $\sim 10$~deg$^{2}$ to contain a few contaminants. However, we note that the depth of our search is valid for detecting kilonovae out to a luminosity distance of \apx100 Mpc. Detecting kilonovae out to the aLIGO BNS horizon at design sensitivity ($\sim200$ Mpc) would require our search to go a magnitude deeper, increasing the expected contamination by a factor of four if we assume simple Euclidean volume scaling and a uniform distribution of sources. 

The challenge of contamination in deep follow-up of GW triggers is significant, but not insurmountable. If localization regions can be reduced to \apx100--200 deg$^2$, we would expect to find \apx170--340 contaminants during a typical search, before making any color cuts. The most aggressive cut, looking for red and non-nuclear sources, would reveal \apx11--22 contaminants in \apx100--200 deg$^2$ localization regions. Obtaining follow-up photometry and NIR spectroscopy for this number of sources, at magnitudes of $z \gtrsim 21$~mag, is a difficult task requiring the allocation of dedicated time on 8-m class telescopes. Looking ahead to the era of $\lesssim 10$ deg$^2$, even our broadest selection criteria will yield fewer than $\sim 10$ candidates. In this regime, obtaining rapid NIR spectroscopic follow-up of sources to assess their true nature becomes tractable and we will truly enter the next generation of multi-messenger astronomy.

{\acknowledgements 
The Berger Time-Domain Group at Harvard is supported in part by the NSF through grants AST-1411763 and AST-1714498, and by NASA through grants NNX15AE50G and NNX16AC22G. P.S.C. is grateful for support provided by the NSF
through the Graduate Research Fellowship Program, grant DGE1144152. The UCSC group is supported in part by NSF grant AST--1518052, the Gordon \& Betty Moore Foundation, and from fellowships from the Alfred P.\ Sloan Foundation and the David and Lucile Packard Foundation to R.J.F. 

This work has made use of data from the European Space Agency (ESA)
mission {\it Gaia} (\url{https://www.cosmos.esa.int/gaia}), processed by
the {\it Gaia} Data Processing and Analysis Consortium (DPAC,
\url{https://www.cosmos.esa.int/web/gaia/dpac/consortium}). Funding
for the DPAC has been provided by national institutions, in particular
the institutions participating in the {\it Gaia} Multilateral Agreement.
This publication makes use of data products from the Two Micron All Sky Survey, 
which is a joint project of the University of Massachusetts and the Infrared Processing 
and Analysis Center/California Institute of Technology, funded by the National Aeronautics 
and Space Administration and the National Science Foundation. This publication makes use 
of data products from the Wide-field Infrared Survey Explorer, which is a joint project of the 
University of California, Los Angeles, and the Jet Propulsion Laboratory/California Institute of 
Technology, funded by the National Aeronautics and Space Administration.
}

\end{document}